\documentclass[9pt]{article}
\usepackage{amssymb}
\usepackage{graphicx}
\begin{document}

\begin{center}
{\Large\bf Power spectrum for the Bose-Einstein condensate dark matter}
\end{center}

\begin{center}
Hermano Velten$^{{\small a,b}}$ and Etienne Wamba$^{c}$

\small{a} Departamento de F\'isica, UFES, Vit\'oria, 29075-910, Esp\'irito Santo, Brazil

{\small b} Fakult$\ddot{\rm a}$t f$\ddot{\rm u}$r Physik, Universit$\ddot{\rm a}$t Bielefeld, Postfach 100131, 33501 Bielefeld, Germany

{\small c}  Laboratory of Mechanics, Department of Physics, Faculty of Science, University of Yaound\'e I, P.O. Box 812, Yaound\'e, Cameroon

\end{center}

\begin{abstract}
We assume that dark matter is composed of scalar particles that form a Bose-Einstein condensate (BEC) at some point during the cosmic evolution. Afterwards, cold dark matter is in the form of a condensate and behaves slightly different from the standard dark matter component. We study the large scale perturbative dynamics of the BEC dark matter in a model where this component coexists with baryonic matter and cosmological constant. The perturbative dynamics is studied using neo-Newtonian cosmology (where the pressure is dynamically relevant for the homogeneous and isotropic background) which is assumed to be correct for small values of the sound speed. We show that BEC dark matter effects can be seen in the matter power spectrum if the mass of the condensate particle lies in the range $15 {\rm meV} < m_{\chi} < 700 {\rm meV}$ leading to a small, but perceptible, excess of power at large scales.

\end{abstract}

\section{Introduction}

Standard cosmology relies on the assumption that dark matter (DM) represents around $25\%$ of the cosmic energy budget and behaves as a pressureless component. This behavior allows the gravitational clustering of the DM particles (once they are decoupled from relativistic species at the early Universe) in order to form dark halos that host galaxies. Such large scale structure supports the hypothesis that DM is composed of weakly interacting massive particles (WIMPs) \cite{DM} though the real nature of this component is still unknown. However, the WIMP scenario faces several challenges at galactic scales \cite{DMproblem}.

The real nature of the dark matter is unknown, i.e. we do not know what it consists of. One possibility is the existence of dark scalar particles (i.e. a spin-$0$ bosons). In this case, the scalar field $\phi$ associated with such particles has its dynamics governed by the potential $V(\phi)$ which encloses all the interactions of such system. Hence, if dark matter particles are bosons they can form, at some critical redshift $z_{cr}$, a Bose-Einstein condensate (BEC), i.e. ground state interacting bosons trapped by an external potential \cite{BECdefinition}. Such BEC stage of the dark matter was proposed in Refs. \cite{Si,JiSi} and it has been widely studied in the literature \cite{BEC} (see also \cite{BECmore}).

In condensed matter physics the mean-field approximation is widely used. In such case, BEC systems are studied through the time-dependent generalized Gross-Pitaevskii equation \cite{pita}
\begin{equation}
i\hbar\frac{\partial \Psi}{\partial t}=-\frac{\hbar^{2}}{2m_{\chi}}\nabla^{2}\Psi+V(r,t)\Psi+ g\, (\left|\Psi\right|^2 ) \Psi,
\end{equation}
where $\Psi\equiv\Psi(r,t)$ is the macroscopic wavefunction of the condensate, $m_{\chi}$ is the mass of the particle, $V(r,t)$ is the trapping potential. The non-linearity term reads
\begin{equation}
g(|\Psi|^{2})=u_{0} |\Psi|^{2}+\epsilon|\Psi|^{4},
\end{equation}
where the quadratic term accounts for the two-body interparticle interaction with $u_0=4\pi \hbar^{2} l_{a}/m_{\chi}$. The term $\epsilon$ is strength of the three-body interparticle interaction which is neglected in numerous works on BEC. The quartic term becomes important
only in the case of higher densities \cite{etienne}. In the standard approach one considers the case where $\epsilon=0$. The free parameters are the boson-boson scattering length ($l_{a}$) and the mass of the particle ($m_{\chi}$). 
 
The application of BEC physics in astrophysics has become much more clear in Ref. \cite{boehmer} where the authors have assumed an arbitrary non-linearity term with $\Psi$ being described by the so-called Madelung representation \cite{madelung},
\begin{equation}
\Psi=\sqrt{\rho(r,t)}e^{\frac{i}{\hbar}S(r,t)}
\end{equation}
where $\rho=\left|\Psi\right|^{2}$ is the density of the condensate. Hence, it comes out that BEC dark matter can be described in terms of a non-relativistic Newtonian fluid with barotropic equation of state that obeys the energy and momentum balances. The BEC pressure is given by \cite{boehmer}
\begin{equation}
P_{be}= g(\rho) \,\rho - \int g(\rho) d\rho.
\end{equation}

In the standard approach of BEC systems $g=u_0 \left|\Psi\right|^{2}$, the effective pressure of the condensate becomes $P_{be}\sim \rho^{2}$. Within this approach remarkable results concerning galaxy curve rotation \cite{boehmer}, dynamics of galaxy clusters \cite{ap1}, the core-cusp problem \cite{harkocorecusp} and galactic vortices \cite{vortices} were obtained (see also \cite{smallBEC}). 

An important issue that should be addressed is whether the background evolution and the large scale structure of the Universe can be affected by the existence of such BEC phase of the dark matter. In reference \cite{harko1} it was assumed that the BEC takes place via a first order phase transition once the temperature of the bose gas reaches the critical value $T_{cr}$. The cosmological parameters at the moment in which such event occurs (i.e. density of the bose gas, the temperature and redshift) were also established in terms of the free parameters of the BEC, namely the mass of the dark matter particle $m_{\chi}$ and $l_{a}$. 

The growth of BEC dark matter perturbations was also considered in references \cite{harko2,chavanis,takeshi}. The general conclusion is that assuming a positive scattering length (that means a repulsive self-interaction) the growth of BEC dark matter inhomogeneities is faster than the standard CDM case ($l_a=0$). This would lead to an excess of power at small scales though the bottom-up structure formation scenario still works. In principle, such excess drives the large scale in a different way resulting in a different power spectrum $P(k)$. As an example, ultra light bosonic DM models ($m_{\chi}\sim 10^{-22}{\rm eV}$) are able to leave very small imprints into the acoustic peaks of the CMB \cite{BECCMB}. However, it has been claimed that although a different small scale clustering the collapsed BEC structures (bosons stars, black holes, etc) behave as cold dark matter at large scales without modifying the large scale power spectrum \cite{takeshi}.

The purpose of this paper is to investigate more carefully the consequences of a BEC dark matter cosmology for the matter power spectrum. Does a Bose-Einstein condensate leave imprints into the $P(k)$? Using the 2dFGRS power spectrum data \cite{cole} we will be able to investigate scales $0.01 Mpc^{-1} \leq k \,h^{-1} \leq 0.185 Mpc^{-1}$ that still belong to the linear perturbation theory. We will assume a cosmological background evolution as proposed in \cite{harko1}, where BEC dark matter coexists with baryons and cosmological constant. We reproduce such model in section 2.1. In section 2.2 we introduce the cosmological perturbation theory using the neo-Newtonian equations. In order to compute the matter power spectrum, perturbations in both dark matter and baryons have to be taken into account. Indeed, the observable power spectrum is the statistical distribution of the visible matter (baryons) which has been driven by the dark matter gravitational field. This step has not yet been done for the BEC scenario. In section 3 we compare the BEC matter power spectrum with the data assuming different model parameters. We conclude in section 4. Throughout this paper, we shall assume $c=1$.
   
\section{Cosmological dynamics of Bose-Einstein condensates}
\subsection{The background evolution}
After the BEC forms its effective pressure can assume a polytropic equation of state such as $P_{be}\sim \rho^{\gamma}_{be}$ if one assumes an arbitrary non-linearity term . The exact value of $\gamma$ is defined by the non-linear contribution of the Gross-Pitaevskii equation which in its standard form leads to \cite{boehmer}
\begin{equation}
P_{be}=\frac{2\pi \hbar^{2} l_a }{m^{3}_{\chi}}\rho^{2}_{be}.
\end{equation}
The scattering length ($l_a$) and the mass ($m_{\chi}$) of the dark matter particle determine the dynamics of the fluid. Assuming that the condensate does not interact with any other form of energy the above pressure leads, via the conservation balance, to
\begin{equation}
\rho_{be}=\frac{m^{3}_{\chi}}{2\pi \hbar^{2} l_a}\frac{\rho_0}{a^{3}-\rho_0}\hspace{1cm} {\rm where} \hspace{1cm} \rho_{0}=\frac{1.266\times\Omega_{be0}\times(m_{\chi}/1{\rm meV})^{-3}\times(l_a/10^9 {\rm fm})}{1+1.266\times\Omega_{be0}\times(m_{\chi}/1{\rm meV})^{-3}\times(l_a/10^9 {\rm fm})}.
\end{equation}
The current value of the scale factor $a$ is $a_0=1$ and the current fractional density of the BEC dark matter is $\Omega_{be0}=\rho_{be0}/\rho_{c}$ where $\rho_{c}$ the critical density. Combining the above relations we obtain the equation of state parameter of the BEC dark matter
\begin{equation}
w_{be}=\frac{\rho_0}{a^3-\rho_0}.
\end{equation} 

A crucial quantity in this model is the moment $z_{cr}$ at which the condensation takes place (i.e. the transition from the ``normal'' DM phase to the Bose-Einstein state). Note that before the transition the bosonic DM particles have decoupled from the primordial plasma and have formed an isotropic gas in thermal equilibrium. From kinetic theory the pressure of a non-relativistic gas in this regime is given by \cite{kinetictheory,harko1}
\begin{equation}
P_{\chi}=\frac{g}{3 h^3}\int \frac{p^2 c^2}{E} f(p)d^3 p\approx 4 \pi\frac{g}{3 h^3}\int \frac{p^4}{m_{\chi}} dp \rightarrow P_{\chi}= \rho_{\chi} c^2 \sigma^{2},
\end{equation}
where $g$ is the number of spin degrees of freedom, $h$ is the Planck constant, $p$ is the momentum of the particle that has energy $E=\sqrt{p^2+m_{\chi}^2 c^4}$ with distribution function $f$. For the velocity dispersion $\sigma^{2}=\left\langle \vec{v}^{2}\right\rangle/3c^{2}$ we assume a mean velocity square $\left\langle \vec{v}^{2}\right\rangle=81 \times 10^{14} cm^2/s^2$, leading to $\sigma^{2}=3\times 10^{-6}$ \cite{harko1}. By imposing the continuity of the pressure at $z_{cr}$ as the thermodynamical condition to be satisfied at the critical redshift, we obtain \cite{harko1}
\begin{equation}
1+z_{cr}=\left(\frac{\rho_0}{\sigma^{2}\left(1-\rho_0\right)}\right)^{-\frac{1}{3\left(1+\sigma^{2}\right)}}.
\label{transition}
\end{equation}
The quantity $\sigma^{2}$ plays the role of the dark matter equation of state parameter for $z>z_{cr}$. However, the usual approach in cosmology considers ``normal'' DM as a standard pressureless fluid.

Since we will assume positive scattering lengths ($0<\rho_{0}<1$) the equation of state parameter $w_{be}$ can be negative if $a^{3}<\rho_{0}$. This can occur in the past. However, typical values of the free parameters produces $\rho \lesssim 10^{-7}$. This means that $w_{be}<0$ for $z \gtrsim{215}$. On the other hand, $\rho \lesssim 10^{-7}$ also implies that $z_{cr}\sim 2$. This means that once the condensate takes place its equation of state is always positive.

Assuming that the transition occurs during the matter dominated phase the radiation fluid can be ignored. The cosmic background expansion $H=\dot{a}/a$ after $z_{cr}$ is given by
\begin{equation}
\frac{H^{2}}{H^{2}_{0}}=\left[\frac{\Omega_{b0}}{a^3}+\frac{\Omega_{be0}(1-\rho_0)}{a^3-\rho_0}+\Omega_{\Lambda}\right],
\label{H}
\end{equation}
where $H_{0}$ is the Hubble constant. Assuming the WMAP7 results \cite{wmap}, the current fractional density of the baryonic component is $\Omega_{b0}=0.045$ and the density parameter of the cosmological constant is $\Omega_{\Lambda}=0.73$. We will assume that the transition occurs suddenly at the redshift $z_{cr}$. This means that the effective expansion before such time is governed by the standard cosmology where $\rho_0=0$. Actually, this assumption is not exactly true. The first order transition occurs at a fixed temperature but it takes some time $\Delta t$ to fully convert normal dark matter into the BEC state. However, as shown in \cite{harko1} for typical values of the BEC parameters the full conversion takes $\Delta t \sim 10^{-4} Gyrs$. This value can be relevant at high redshifts but as we will show, concerning the matter power spectrum, the present model displays more appreciable features if the transition happens at low redshift where $\Delta t \sim 10^{-4} Gyrs$ is  negligible.

\subsection{Perturbations using the neo-Newtonian cosmology}

At scales larger than the horizon the Newtonian theory fails and the full relativistic equations should be adopted. However, cosmology can be understood within the Newtonian framework if we properly interpret the physical quantities like velocity, energy density and gravitational potential. Such approach is known as Newtonian cosmology that is a useful aproximation for a Einstein-de Sitter Universe. However, if the inertial effects of the pressure become relevant as for example during the radiation phase or at the onset of the accelerated expansion, Newtonian Cosmology also fails and we need a more appropriate set of equations. The inclusion of pressure in the Newtonian cosmology in order to make it relevant for the homogeneous and isotropic background gave rise to the neo-Newtonian cosmology \cite{NeoNewtonian}. 

A relativistic component sources the gravitational field through the energy-momentum tensor
\begin{equation}
T^{\mu\nu}=(\rho +P)u^{\mu}u^{\nu} + P g^{\mu\nu}.
\end{equation}
Contracting $T^{\mu\nu}$ with $u_{\mu}$ and $h_{\mu\alpha}=g_{\mu\alpha}+u_{\mu}u_{\alpha}$ we obtain, respectively
\begin{equation}
\dot{\rho}+\nabla_{r} (\rho \vec{v})+P\nabla_r \vec{v}=0,
\label{continuity}
\end{equation}
\begin{equation}
\dot{\vec{v}}+(\vec{v}.\nabla_{r}) \vec{v}=-\nabla_r \phi-\frac{\nabla_r P}{\rho+P}-\frac{\dot{P}\vec{v}}{\rho+P}.
\label{euler}
\end{equation}
Since in the neo-Newtonian cosmology it is assumed that the effective energy of the fluid is the trace of $T^{\mu\nu}$, then the pressure comes into play. Moreover, the gravitational interaction occurs via the modified Poisson equation
\begin{equation}
\nabla^{2}\phi=4\pi G (\rho+3P).
\label{poisson}
\end{equation}
 Equations (\ref{continuity}-\ref{poisson}) were used in \cite{harko2} in order to study cosmological perturbations. However, it is expedient to expose a cautionary remark. The neo-Newtonian first-order perturbation dynamics and its relativistic counterpart coincide in the case
of a vanishing sound speed only \cite{reis}. Hence, one may expect that the correct relativistic results are reproduced by the neo-Newtonian perturbation
dynamics on all perturbation scales at least for small values of the sound speed.

The cosmological constant does not fluctuate in the standard approach. Both the BEC dark matter and the baryons obey separately the equations (\ref{continuity}) and (\ref{euler}). The r.h.s of the Poisson equation will include the contributions from all components. Our set of equations will be composed of five equations. We introduce perturbations in such equations writing each quantity $f=\left\{\rho,\vec{v},P,\phi\right\}$ as $f\rightarrow f+\hat{f}(\vec{r},t)$ where the symbol  ``\,\^\,'' (hat) means a first order quantity. This allows to calculate the matter density contrast $\delta\equiv \hat{\rho}/\rho$ that will be used to compute the power spectrum.

Collecting the first order terms and Fourier transforming the perturbations as $\hat{f}(\vec{r},t)=\delta f(t) e^{\frac{i \vec{k}.\vec{r}}{a}}$, with $k$ being the wavenumber of the perturbation, we end up with (details of such calculations can be found in \cite{neoCha})

\begin{equation}\label{deltab}
\delta^{\prime\prime}_{b}+\delta_{b}^{\prime}\left(\frac{H^{\prime}}{H}+\frac{3}{a}\right)-\frac{3}{2}\frac{\Omega_{b}}{H^{2}a^{2}}\delta_{b}=\frac{3\Omega_{be}}{2H^{2}a^{2}}(1+c^{2}_{s})\delta_{be},
\end{equation}

\begin{eqnarray}\label{deltabec}
\delta^{\prime\prime}_{be}+\left(\frac{H^{\prime}}{H}+\frac{3}{a}-\frac{w^{\prime}_{be}}{1+w_{be}}-\frac{3w_{be}}{a}\right)\delta_{be}^{\prime} + \\ \nonumber
\left\{3w_{be}\left[\frac{H^{\prime}}{Ha}+\frac{(2-3w_{be})}{a^2}\right]+\frac{3w_{be}^{\prime}}{a(1+w_{be})}+\frac{\left(k/k_0\right)^{2}c^{2}_{s}}{H^{2}a^{4}}-\frac{3}{2}\frac{\Omega_{be}}{H^2a^2}(1+3c^{2}_{s})(1+w_{be})\right\}\delta_{be}
=\frac{3}{2}\frac{\Omega_{b}}{H^{2}a^{2}}(1+c^{2}_{s})\delta_{b},
\end{eqnarray}
where $k_{0}^{-1}=3000 h Mpc$ is the present Hubble length. In equations (\ref{deltab}) and (\ref{deltabec}) the symbol $\prime$ means derivative w.r.t. the scale factor. The density contrast of the baryonic fluid and the BEC dark matter are, respectively $\delta_{b}$ and $\delta_{be}$. The speed of sound $c^{2}_{s}$ of the condensate fluid is 
\begin{equation}
c^{2}_{s}=\frac{\partial \rho_{be}}{\partial P_{be}}=2w_{be},
\end{equation}
which is a fundamental quantity for the power spectrum.
If the BEC behaves as standard cold dark matter, equations (\ref{deltab}) and (\ref{deltabec}) admit the usual solution $\delta_{b}\sim\delta_{be}\sim a$. In the next section the above set of equations will be solved for different values of the parameters $l_{a}$ and $m_{\chi}$.

\section{The matter power spectrum}

The matter (baryonic) power spectrum is defined as 
\begin{equation}
P(k)=\left|\delta_{b}(z=0;k)\right|^{2},
\end{equation}
where $\delta_{b}(k)$ is baryonic density contrast calculated in equations (\ref{deltab}) and (\ref{deltabec}) at the present time. The baryonic agglomeration $\delta_{b}$ is driven by the gravitational field which is sourced by all the forms of energy.

In order to solve equations (\ref{deltab}) and (\ref{deltabec}) we need to set the initial conditions for $\delta_{b}$ and $\delta_{be}$ and their derivatives at $z_{cr}$ where the condensate takes place. Since $z_{cr}=z_{cr}(l_{a},m_{\chi})$ for each chosen couple of  values $(l_{a},m_{\chi})$ we need different initial conditions. 

Assuming a primordial Harrison-Zeldovich power spectrum $P\sim k$ and evolving it with the appropriate growth function and the BBKS transfer function \cite{bbks} we obtain the theoretical expression for the power spectrum today. We will denote such result by $P_{HZ}(z)$. As the $\Lambda$CDM model also fits the data we then integrate back in time the perturbed $\Lambda$CDM equations from $z=0$ (where the spectrum $P_{HZ}(z)$ is assumed) until $z=z_{cr}$. With this approach we have the standard power spectrum at $z_{cr}$ that will be used as initial conditions for the BEC model. Now, we are able to evolve equations (\ref{deltab}) and (\ref{deltabec}) from $z_{cr}$ until today. 

Since the linear scales probed by the 2dFGRS \cite{cole} data correspond to $k/h < 0.185 Mpc^{-1}$, the final spectrum for the BEC cosmology is normalized in such way that $P_{BEC}(k \, h^{-1}=0.185 Mpc^{-1})=P_{HZ}(k\, h^{-1}=0.185 Mpc^{-1})$.

\begin{figure}[!t]
\begin{center}
\includegraphics[width=0.47\textwidth]{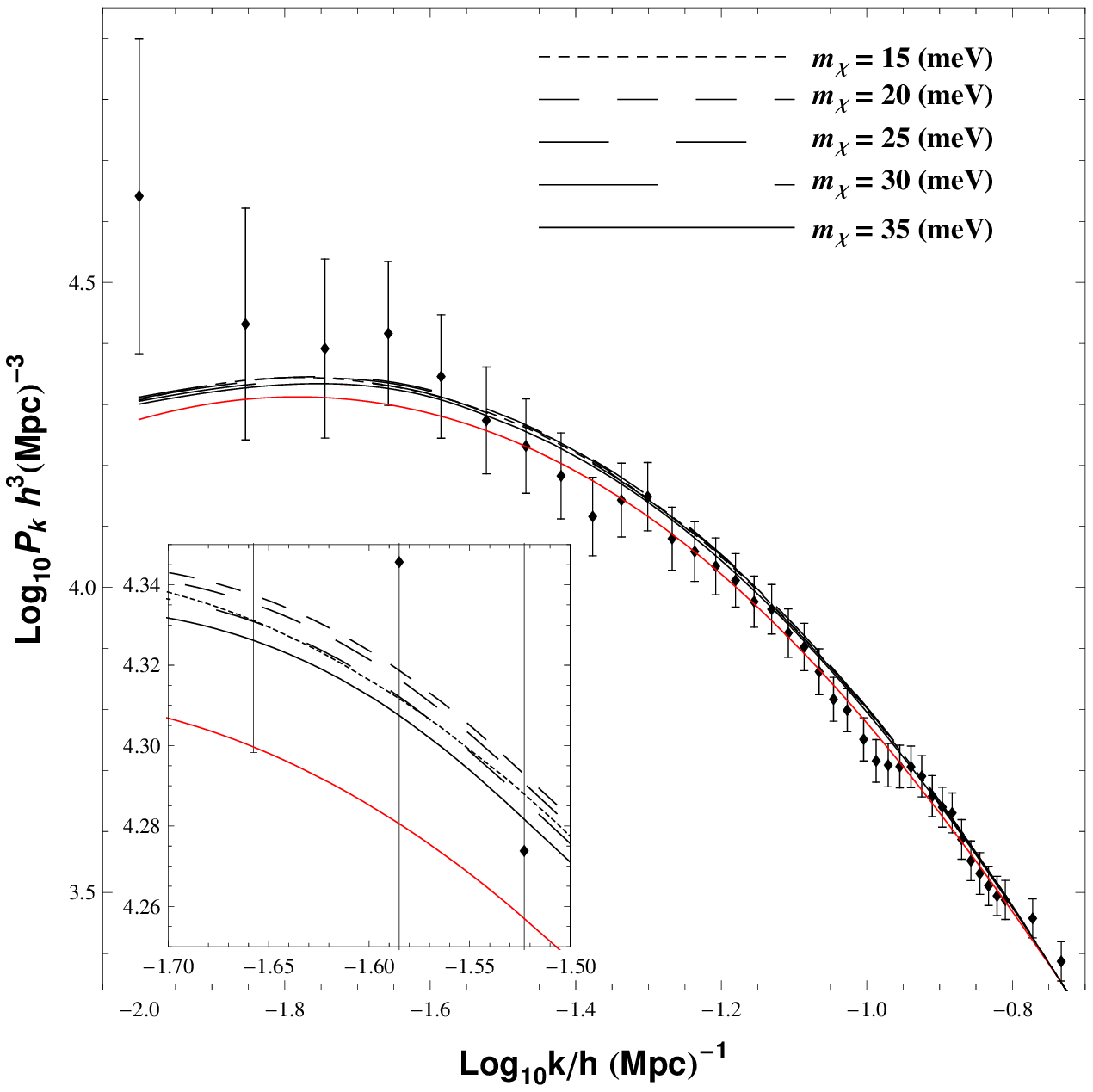}
\includegraphics[width=0.49\textwidth]{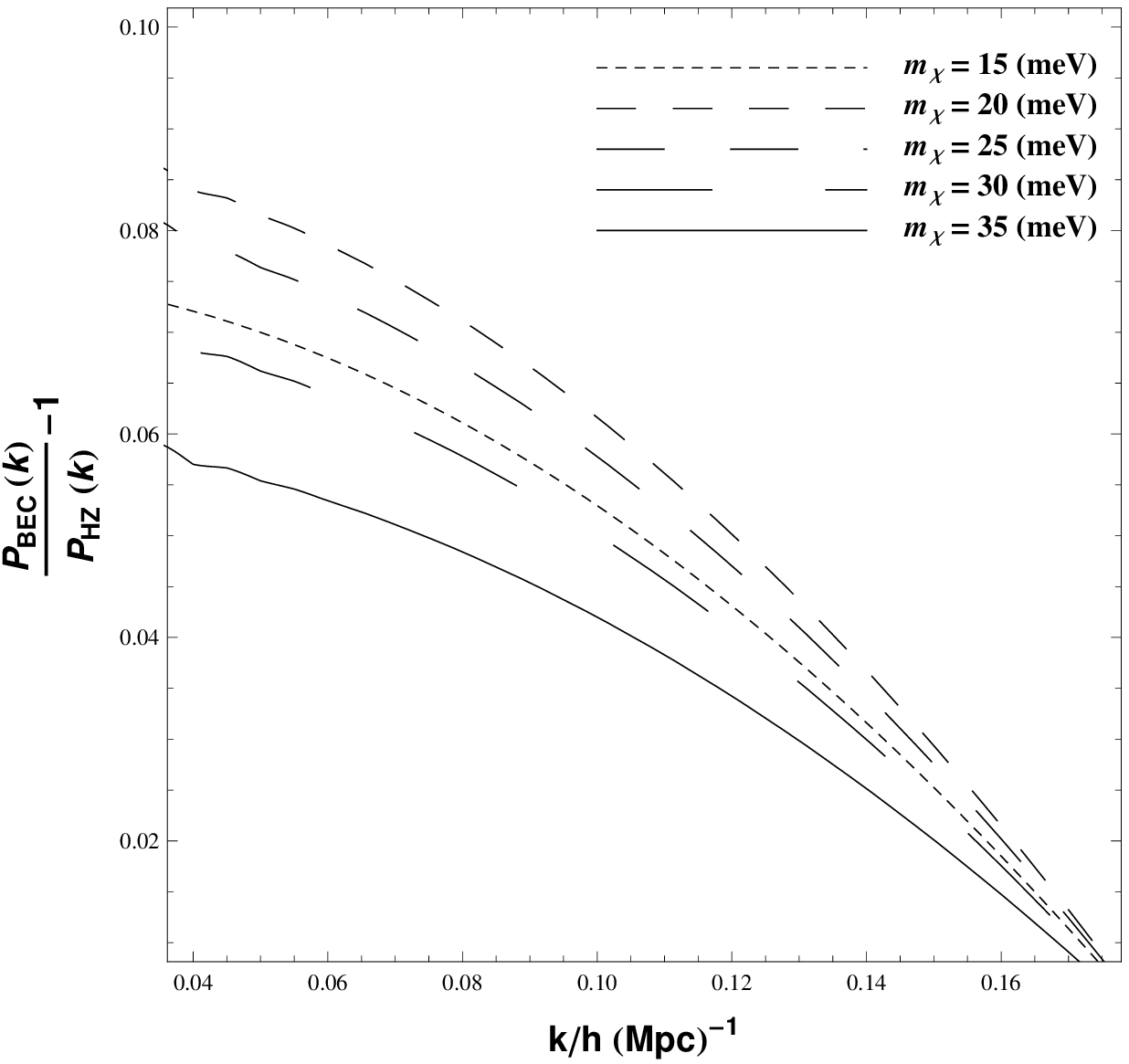}
\includegraphics[width=0.47\textwidth]{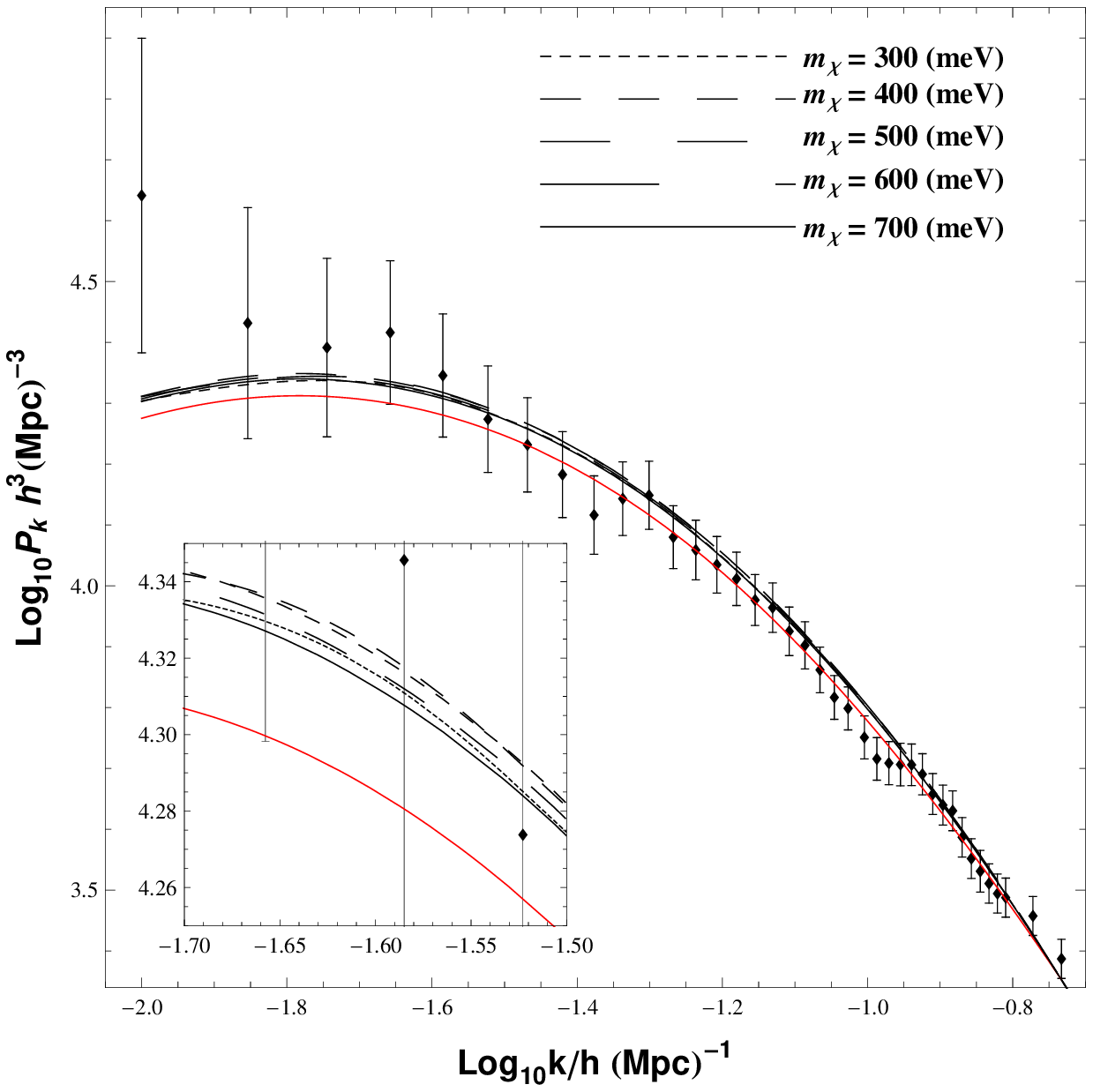}
\includegraphics[width=0.49\textwidth]{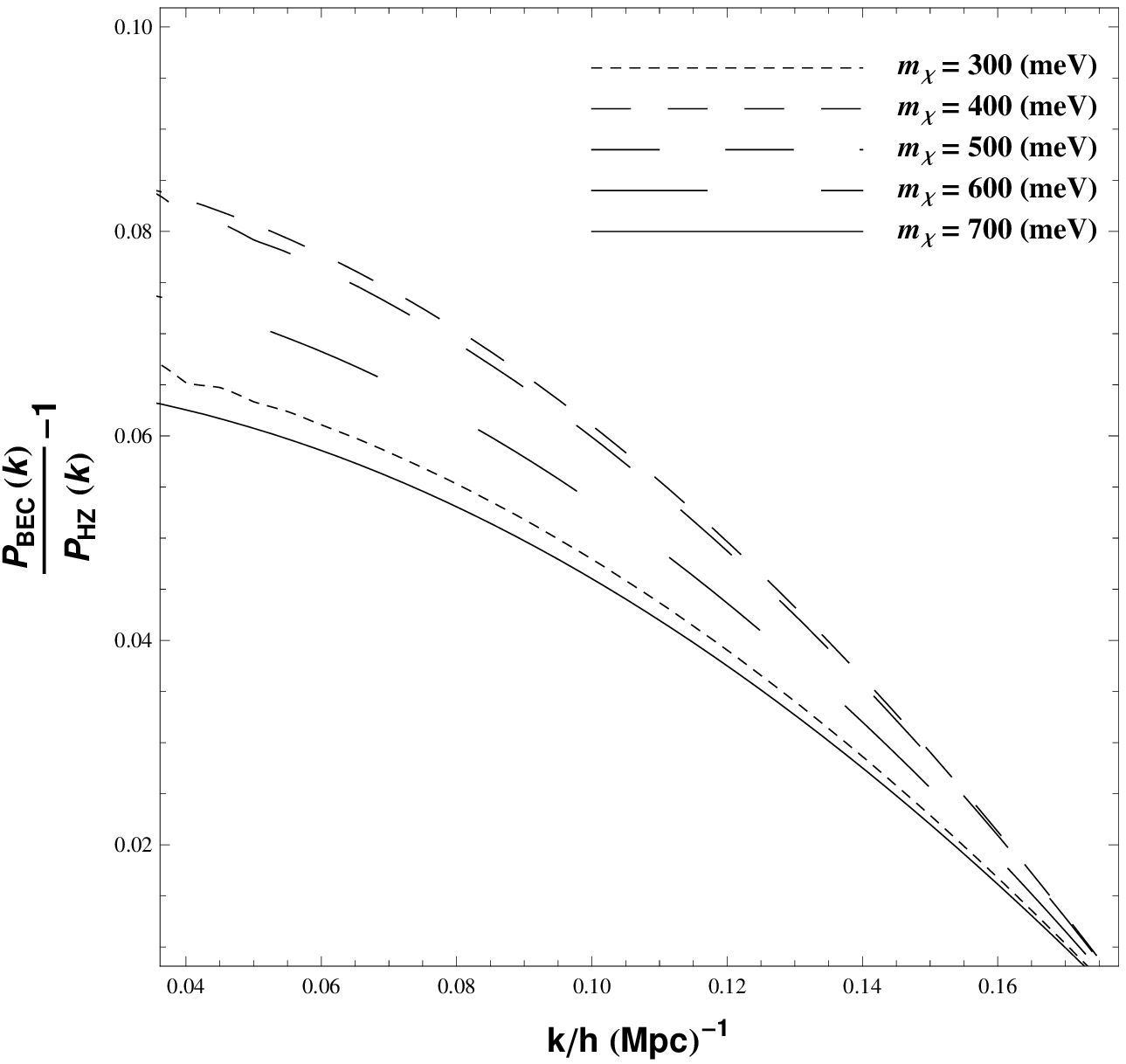}
\label{f1}	
\caption{Matter Power Spectrum. Red line is the Harrison-Zeldovich spectrum with BBKS transfer function. BEC cosmology with $l_{a}=10^{6}{\rm fm}$ ($l_{a}=10^{10}{\rm fm}$) was adopted with different masses $m_{\chi}$ in the top-left (bottom-left) panel. Top-right (bottom-right) panel shows the relative difference between BEC models with $l_a={10^{6}}{\rm fm}$ ($l_{a}=10^{10}{\rm fm}$) and the standard Harrison-Zeldovich power spectrum for various masses $m_{\chi}$.}
\end{center}
\end{figure}

In the left panels of figure 1 we show the power spectrum for some model parameters. The data points come from the 2dFGRS project \cite{cole}. In the top-left (bottom-left) panel we have fixed $l_{a}=10^{6}{\rm fm}$ ($l_{a}=10^{10}{\rm fm}$) and calculated the power spectrum for the BEC model $P_{BEC}(k)$ assuming different values of the mass $m_{\chi}$. These values adopted here for $l_{a}$ are the usual scattering lengths found in laboratory. The red solid line in both panels corresponds to the power spectrum $P_{HZ}(k)$ obtained from the BBKS transfer function. The right panels show the relative difference $\left(P_{BEC}\left(k\right)-P_{HZ}\left(k\right)\right)/P_{HZ}\left(k\right)$. 

A remarkable point of the present model is that if $l_{a}=10^{6}{\rm fm}$ ($l_{a}=10^{10}{\rm fm}$) the transition occurs in the future, i.e. $z<0$, for masses $m_{\chi}<5{\rm meV}$ ($m_{\chi}<100{\rm meV}$). This sets a lower bound to $m_{\chi}$ that can be tested using current observations.

The BEC cosmology exhibits an excess of power for any values of the model parameters. The chosen masses correspond to the largest deviations from $P_{HZ}(z)$ we have found. Although the small differences they are indeed perceptible. For example, if $l_{a}=10^{6}{\rm fm}$, a typical value for the scattering length, the difference can achieve $\sim 8\%$ if $m_{\chi}=20 {\rm meV}$ (see top-right panel in figure 1). Deviations of the same order have also been found for a mass $m_{\chi}=400 {\rm meV}$ if $l_{a}=10^{10}{\rm fm}$ which is the largest scattering length used in laboratory. For the parameters values used in figure 1 we show the corresponding critical redshift $z_{cr}$ in table 1. Note that the masses adopted here obey the constraint $m_{\chi} < 19 {\rm eV}$ found in Ref. \cite{takeshi}.

\begin{table}[!h]
\begin{center}
 \begin{tabular}{|c|c|c|c|c|c|c|c|c|c|c|c|}
 \hline
  & \multicolumn{5}{|c|}{$l_{a}=10^{6}{\rm fm}$}& \multicolumn{5}{|c|}{$l_{a}=10^{10}{\rm fm}$}  \\ 
      \hline  
      $m_{\chi} ({\rm meV})$     & $15$   & $20$   & $25$  & $30$  & $35$  & $300$   & $400$   & $500$  & $600$  & $700$  \\ \hline      
      $z_{cr}$      & 2.17  & 3.23    & 4.29  & 5.35 & 6.40 & 1.95  & 2.93  & 3.91 & 4.89 & 5.88 \\ \hline       
   \end{tabular}
      \end{center}
\caption{Values of the masses $m_{\chi}$ used in Figure 1 with the corresponding critical redshift $z_{cr}$ at which the transition to the BEC phase occurs. }
\end{table}

\section{Conclusions}

We have assumed that dark matter is composed of scalar particles that are able to form a Bose-Einstein condensate at some critical redshift $z_{cr}$. At this point, a first order phase transition drives the conversion of ``normal'' dark matter into the BEC. After $z_{cr}$ the dynamics of the dark matter component evolves in a slightly different way \cite{harko1}. Consequently, one can also expect a different perturbative dynamics. Indeed BEC dark matter accelerates the gravitational clustering at small scales \cite{harko2,chavanis} but there are claims that at large scales the BEC dark matter behaves effectively as a typical cold dark matter component \cite{takeshi}. 

Using the matter power spectrum we have shown that if such phase transition has occurred in the recent Universe this process was able to leave small, but perceptible, imprints on the large scale structure. Assuming $l_{a}=10^{6} {\rm fm}$ the BEC dark matter models shows corrections of the order of a few percents for masses $15-35 \,{\rm meV}$. Adopting $l_{a}=10^{10} \,{\rm fm}$ corrections of the same order are obtained for masses $300-700 {\rm meV}$. 

For the relevant parameter values studied here (see table 1) the transition to the BEC phase occurs at low redshifts. Since the standard cosmology remains unchanged before $z_{cr}$ the CMB physics at the last scattering surface will be the same. However, the BEC dark matter would modify the gravitational potential just after $z_{cr}$ while the speed of sound is nonzero leading to a contribution to the integrated Sachs-Wolfe effect. Such analysis needs the relativistic perturbation theory which is beyond the scope of this work. 

Although the small influence of the BEC phase on the matter power spectrum a more quantitative analysis can be performed using Bayesian statistics. This would estimate the preferred values of the model parameters. We leave this analysis for a future work. Also, for a more general study of BEC systems we can consider the case $\epsilon \neq 0$ which would provide a richer dynamics for the BEC system.

{\bf Acknowledgments}

HV is supported by Conselho Nacional de Desenvolvimento Cient\'ifico e Tecnol\'ogico (CNPq-Brazil). EW is thankful to the
organizers of the Benin Second Afro-Brazilian Meeting of Physics and
Mathematics supported by Programa pr\'o-\'Africa (Capes-Brazil) and ICTP. We would like to thank J\'ulio Fabris
for helpful comments on this paper.


\begin{thebibliography}{99}
\bibitem{DM} Gianfranco Bertone, Dan Hooper and Joseph Silk, Physics Reports {\bf405}, 279 (2005). 
\bibitem{DMproblem} J. Diemand, B. Moore, and J. Stadel, Nature, 433, 389 (2005); J. Diemand, M. Kuhlen, and P. Madau, Astrophys. J.
657, 262 (2007).
\bibitem{BECdefinition} L. Pitaevskii and S. Stringari, {\it Bose-Einstein Condensation} (Clarendon, Oxford, 2003).
\bibitem{Si} S. J. Si, Physical Review {\bf D50}, 3650 (1994).
\bibitem{JiSi} U. S. Ji and S. J. Si, Physical Review {\bf D50}, 3655 (1994).
\bibitem{BEC} W. Hu, R. Barkana and A. Gruzinov, Physical Review Letters {\bf 85}, 1158 (2000); S. Khlebnikov, Physical Review {\bf A 66}, 063606 (2002); James E. Lidsey, Class.Quant.Grav. {\bf21}, 777 (2004); F. Ferrer and J. A. Grifols, JCAP 0412 (2004) 012.
\bibitem{BECmore} F. Siddhartha Guzman and L. Arturo Ure\~na-L\'opez, Astrophys.J.{\bf 645}, 814 (2006); L. Arturo Ure\~na-L\'opez, JCAP 0901:014 (2009); L. Arturo Ure\~na-L\'opez and Argelia Bernal, Phys.Rev.D {\bf82}, 123535 (2010).   
\bibitem{pita} E. P. Gross, Nuovo Cimento, {\bf 20}, 454 (1961); L. P. Pitaevskii, Zh Eksp Teor Fiz {\bf 40} 646 (1961)[Sov Phys JETP {\bf 13} 451].
\bibitem{etienne} E. Wamba, A. Mohamadou and T. Kofan\'e, Physical Review {\bf E77}, 046216 (2008); A. Mohamadou, E. Wamba, S. Y. Doka, T. B. Ekogo and T. C. Kofane,  Physical Review {\bf A84}, 023602 (2011).
\bibitem{boehmer} C. Boehmer and T. Harko, JCAP 0706, 025 (2007). 

\bibitem{madelung} E. Madelung, Zeit. F. Phys. {\bf 40}, 322 (1927).
\bibitem{ap1} Jae-Weon Lee at al, arXiv: 0805.3827.
\bibitem{harkocorecusp} T. Harko  JCAP 1105, 022 (2011). 
\bibitem{vortices} Ben Kain and Hong Y. Ling, Physical Review {\bf D82} 064042 (2010). 
\bibitem{smallBEC} Pierre-Henri Chavanis, Physical Review {\bf D84} 043531 (2011); P.H. Chavanis and L. Delfini, Physical Review {\bf D84} 043532 (2011);  
\bibitem{harko1} T. Harko  Physical Review {\bf D83} 123515 (2011). 
\bibitem{harko2} T. Harko, MNRAS {\bf413} 3095 (2011).
\bibitem{chavanis} P. H. Chavanis, arXiv:1103.2698. 
\bibitem{takeshi} T. Fukuyama, M.Morikawa and T. Tatekawa JCAP 0806, 033 (2008); Progress of Theoretical Physics, {\bf 115}, 6 (2006)
\bibitem{BECCMB} I. Rodr\'igues-Montoya et al. The Astrophysical Journal, {\bf721} 1509 (2010).
\bibitem{cole} S. Cole et al., MNRAS, {\bf 362}, 505 (2005).
\bibitem{kinetictheory} C. J. Hogan and J. J. Dalcanton, Phsy, Rev. D {\bf 62}, 063511 (2000); J. Madsen, Phys. Rev. D {\bf 64}, 027301 (2001).


\bibitem{wmap}http://lambda.gsfc.nasa.gov
\bibitem{NeoNewtonian} W.H. McCrea, Proc. R. Soc. London {\bf 206}, 562
(1951); E.R. Harrison, Ann. Phys. (N.Y.) {\bf 35}, 437
(1965); J.A.S. Lima, V. Zanchin and R. Brandenberger,
Month. Not. R. Astron. Soc. {\bf 291}, L1(1997).
\bibitem{reis} R.R.R. Reis, Phys. Rev. {\bf D67}, 087301 (2003); erratum-ibid {\bf D68}, 089901(2003).
\bibitem{neoCha}  J.C. Fabris, S.V.B. Gon\c{c}alves, H.E.S. Velten and W. Zimdahl, Physical Review \textbf{D78}, 103523 (2008).
\bibitem{bbks} J.M. Bardeen, J.R. Bond, N. Kaiser and A.S. Szalay, Astrophys. J., {\bf 304} 15 (1986); J. Martin, A.
Riazuelo and M. Sakellariadou, Physical Review {\bf D61}, 083518 (2000).


\end{thebibliography}
\end{document}